# Characterization of Low-Pressure MWPC from 1E3 to 1E5 Pa

Yanfeng Wang[a, b, c], Zhijia Sun[b, c, d], Xiaohu Wang[a, e], Han Yi[b, c], Wei Jiang[b, c], Ruirui Fan[b, c, d1], Liang Zhou[b, c], Yongcheng He[b, c], Changjun Ning[b, c], Yuefeng He[b, c], Yingtan Zhao[b, c], Kang Sun[b, c], Keqing Gao[b, c]

a    Fundamental Science on Nuclear Wastes and Environmental Safety Laboratory, Sichuan, China
b    Institute of High Energy Physics, Chinese Academy of Sciences (CAS), Beijing 100049, China
c    Spallation Neutron Source Science Center, Dongguan 523803, China
d    State Key Laboratory of Particle Detection and Electronics, Beijing 100049, China
e    Southwest University of Science and Technology, Mianyang 621010, China

## Abstract:

The LPMWPC can be used as the $\Delta E$ detector for the low-energy charged particle detection. In order to increase the transmittance, the wires were adopted as the cathode. This work investigated the LPMWPC signal characteristics of this configuration and measured the gas gain with a mixture of 90%Ar and 10% $CO_2$ from 1E3 to 1E5 Pa. From the test, the second pulse after the avalanche signal was observed, which proved to be caused by the ions' drifting near the cathode wire.

[1] Correspondence to: No.1 Zhongziyuan Road, Dalang, Dongguan 523000, China.
*E-mail addresses*: fanrr@ihep.ac.cn (R. Fan)

## 1 Introduction

In the nuclear data measurements and nuclear physics experiments, the $\Delta E$-$E$ is an effective method for charged particle identification. Usually, a thin silicon was used as the $\Delta E$ detector with a threshold about several MeVs. [1-2] To detect the even lower energy charged particle, the gas-based Ionization Chamber (IC), which has the larger active area and the lower materials, is a solution. [3-4] The signal of the IC is proportional to the deposited energy of the incident particle, which is depended on the thickness and pressure of the working gas. To gain a better Signal-to-Noise Ratio (SNR), the higher gas pressure and the thicker volume are needed. But for the low-energy charged particle detection, a vacuum environment is usually required. The corresponding gas sealing window of the IC is also tens microns, which will stop the low-energy charged particle or cause a lot of energy consumption. The lower pressure IC has a thinner window, but less the energy deposited. Meanwhile, the size of the detector is limited in the vacuum chamber.

The Low-Pressure Multi-Wired Proportional Chamber (LPMWPC) has a large gas gain. It means that the avalanched signal as low as tens of keVs can be obtained in the LPMWPC, which is equivalent to only thousands of primary ionizations. Since 1971, a series of experiments had been performed to investigate the energy and time resolution of LPMWPC. [5-12]

An LPMWPC-based $\Delta E$-$E$ telescope array was developed at Back-n white neutron source. With this telescope array, several experiments were accomplished, and the lowest energy detection was 0.5 MeV for protons and 1.01 MeV for $^7$Li. The pressures of the LPMWPC can be tuned from 1E3 Pa to 1E5 Pa for different production measurements. [13-14]

During the testing, the different characteristics of the LPMWPC was observed. The following paragraphs describe the detector construction, the observed signal and the gas gain testing. Some simple derivations to illustrate this phenomenon are also included.

## 2 The LPMWPC construction

The LPMWPC consists of three same electrodes, two grounded for cathodes (top and bottom), and one for the anode (in the middle). Each electrode is made of the gold-coated tungsten wires with a diameter of 25 μm. The wires are welded in every 2 mm on a hollow PCB which has an empty area of $30 \times 30$ mm$^2$. All wires on a frame are connected and read out together. Nylon spacers keep a gap of 2 mm between each frame. A Si-PIN detector with a thickness of 300 $\mu$m is placed behind the LPMWPC as the total energy detector. The photo of a $\Delta E$-$E$ telescope as shown in Fig. 1. The thin PET-film windows from 0.5 to 6 μm keep the working gas mixture at various pressures. A stainless-steel woven mesh outside of the box with 52% transmittance is employed to support the window in the vacuum chamber. This pressure-tunable design helps the LPMWPC get adequate deposited energy (normally >20 keV) from different (n, lcp) measurements. Meanwhile, it also ensures that most of the charged particles can travel through the LPMWPC and stop in the Si-PIN detector. Because the flammable gas is strictly prohibited in the tunnel, in the test the gas mixture of 90% argon and 10% $CO_2$ was filled in the LPMWPC.

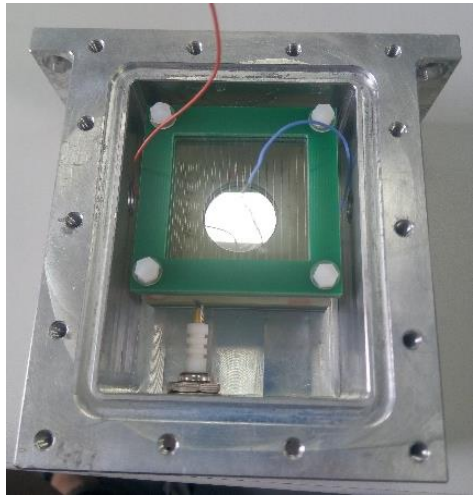

Fig. 1. The photo of the ΔE-E telescope in an aluminum shielding box, a 300 um Si-PIN detector with a diameter of 20 mm followed as the total E detector

## 3 Characteristics of Signals

Followed tests were performed with a $^{241}$Am $\alpha$ source. The LPMWPC is biased through a charge sensitive preamplifier "MPR-1" from the Mesytec GmbH & Co [15]. The output of the preamplifier is digitized by the General-Purpose Readout Electronics (GPRE from USTC). [16] A customized Digital Signal Processing (DSP) network based on the ROOT software handles all digital waveforms from the detector. This DSP contains a simple smoothing, a high pass, and five low-pass filters to make the signal a semi-Gaussian distribution. Fig. 2 shows the waveform before and after the DSP at the pressures of 1E5 (the normal pressure MWPC), 2E4, and 3E3 Pa.

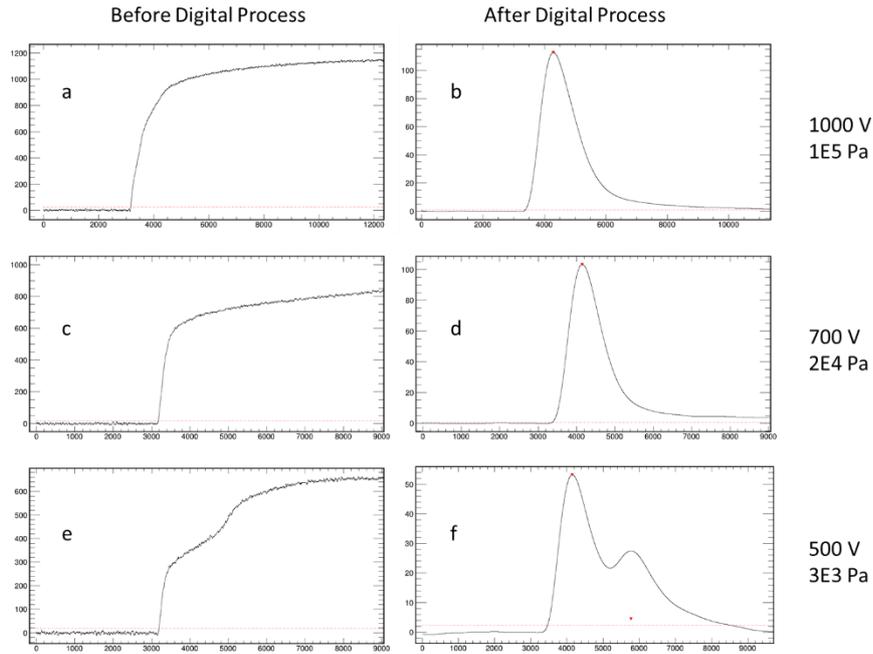

Fig. 2. The MWPC works at different pressures, leading to different signal waveforms. The gas pressures are 1E5, 2E4, and 3E3 Pa from top to bottom. The plots on the left (a, c, e) show the original waveform, while the waveform after the DSP on the right (b, d, f) with the X-axis in nanosecond.

It can be noticed from the original waveforms Fig.2 (a) and (c) that there are slow risings (ion tail) after the signal, which are caused by the ions' slow drifting. The rising turned more significant with the lower pressure, due to the faster velocity of the ions. At even lower pressure (3E3 Pa) shown in Fig.2 (e), a second pulse was observed in microseconds after the first pulse.

This phenomenon has been described in reference [17], and briefly illustrated in Fig. 3. Avalanche occurs near the surface of the anode wire, typically tens or hundreds of microns, which is depended on the electric field and the pressure (as the reduced field $E/p$). While the electrons are collected by the anode wire immediately, the avalanche ions drift away. Both of the charge motions cause the first pulse. While the ions go through the weak field area, an ion tail is induced, especially in a charge sensitive preamplifier. When the ions approaching the vicinity of the cathode wire, and a more rapid signal (the second pulse) is generated due to the higher field. Under lower pressures, the reduce electric field $E/p$ can be much larger than normal, leading a significant ion induced signal.

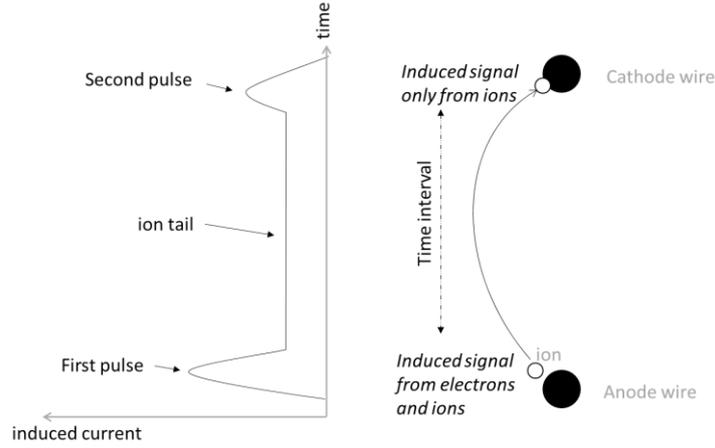

Fig. 3. The schematic of the second pulse in the LPMWPC waveform

Since the ion spends most of time in travelling through the low field area. By simplification similar to the parallel plate electric field, the time interval $\tau$ between the two signals can be approximated as:[18]

$$\tau \approx \frac{d}{v} \approx \frac{dp}{E\mu_0} \approx \frac{d^2 p}{U\mu_0} \qquad (1)$$

Where $d$ is the gap between anode and cathode planes, $\mu_0$ is the mobility of the ions, $U$ is the bias voltage. From waveforms after the PSD, the time interval between peaks can be obtained. For instance, Fig. 4 shows the voltage-dependent waveforms at a constant pressure of 1E4 Pa.

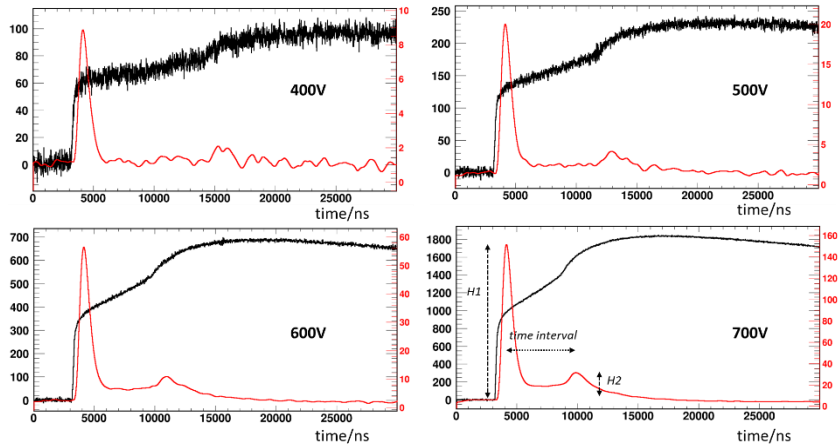

Fig. 4. The waveforms vary from different biases at a constant pressure of 1E4 Pa. In each plot, the black curve is the original pulse, while the red one is after the PSD. The vertical axis is in an arbitrary unit. The horizontal axis is time in nanoseconds.

The experimental velocity of ions in this test has a linear correlation with the $E/p$ as Fig. 5 shown.

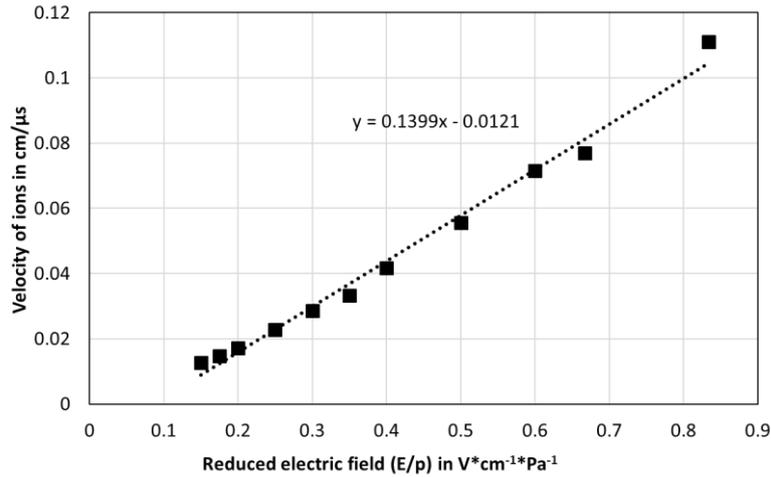

Fig. 5. The velocity of ions versus reduced electric field in the gas mixture of Argon (90%) and $CO_2$ (10%)

First peak height (*H1*) and second peak height (*H2*) in Fig.4 can be measured from the waves. Different from the first pulse, only ions' drifting, especially near the cathode wire, induces the second pulse. The ratio of *H2/H1* represents a ions' rough contribution of the total induced charge. Table. 1 lists some *H2/H1* from the tests with different pressures and biases. *H2/H1* decreases with the reduced electric field.

Table. 1. The H2/H1 and time interval at different pressures and biases

| Pressure/Pa | Bias/V | *H2/H1* | The time interval between *H1* and *H2* /ns |
|---|---|---|---|
| 2.00E+04 | 700 | 0.42 | 13600 |
| 5.00E+03 | 600 | 0.37 | 2800 |
| 3.00E+03 | 500 | 0.32 | 1800 |

From a micro perspective, with a higher reduced electric field, the avalanche starts in the farther range from the anode wire. This leads the electrons induced more charge. At even lower pressure (<1E3 Pa), the electrons' signal becomes the dominant and rapid enough for timing. Therefore, the LPMWPC is often employed as the timing detector before. [5-9]

## 4 Gas gain

To get the primary ionization, a voltage of 50 V was applied on the LPMWPC, in which the detector works in the ionization mode. The relation between signal amplitude and the gas pressure was obtained as in Fig. 6. The 5.41 MeV α particle penetrates the gas chamber, so the energy deposited in LPMWPC has an almost linear relationship with the gas pressure.

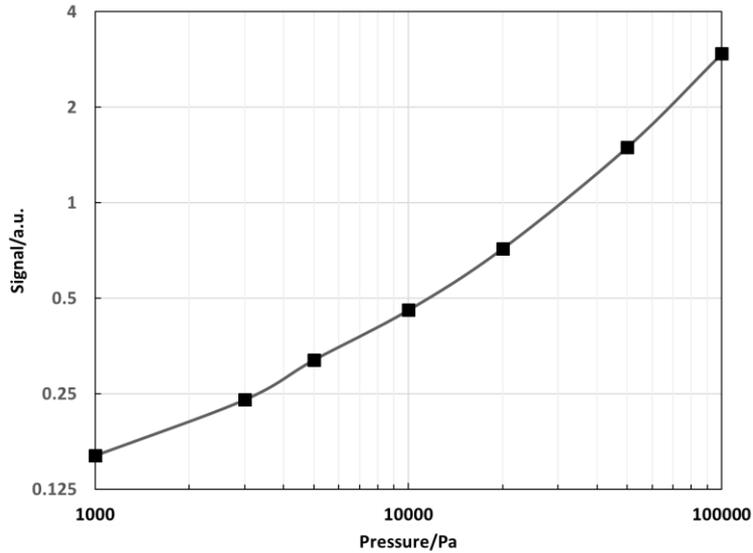

Fig. 6. With 50 V bias, the LPMWPC is working in the ionization mode, and the primary ionizations are not amplified. We obtain a linear relationship between the signal amplitude and pressure, namely the energy deposited is proportional to the gas density.

Fig. 7 demonstrates the normalized gas gain curve at various pressures. The multiplication (gas gain) increases with the bias, the trend follows an approximately exponential function. In this multiplication calculation, only the $H1$ is counted as the signal amplitude. The multiplication slope decreases with the increasing pressure.

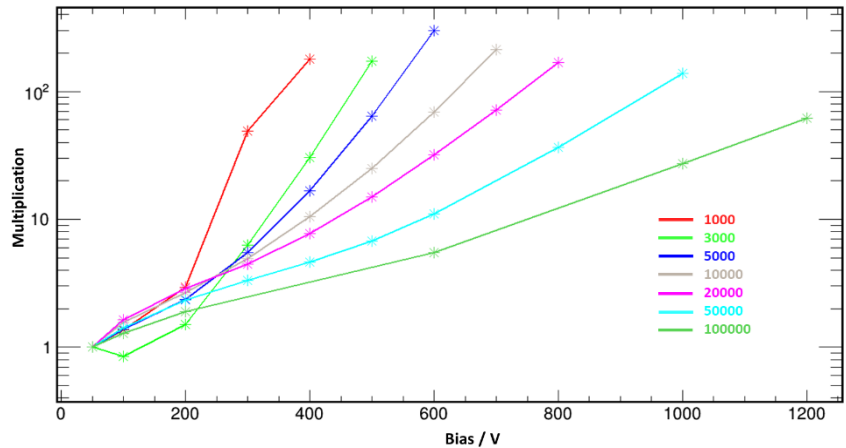

Fig. 7. The gas gain test results from 1E3 to 1E5 Pa. The signal is normalized with the primary ionization. The detector multiplication factor increases with the increasing bias in each pressure test.

## 5 Discussion

In order to minimize the energy loss of incident particles, it is the best to use the wires as the cathode. In this case, the second pulse caused by ions' drifting is the main reason that affects the particle resolution, which might be mistaken for an accumulated physical event. The preliminary test of LPMWPC can help confirm the time interval and the ratio of the second peak to signal. In principle, it improves the particle resolution of the *ΔE-E* detector.

A good example is the results from $^{10}$B (n, alpha) $^7$Li experiment in reference [14], where the LPMWPC worked at 3000 Pa. At this pressure, a significant double pulse is shown as in Fig.8 (a), with a time interval of 2 microseconds. A simple and effective way to solve this problem is to increase the integration time of the PSD, as in the Fig. 8 (b).

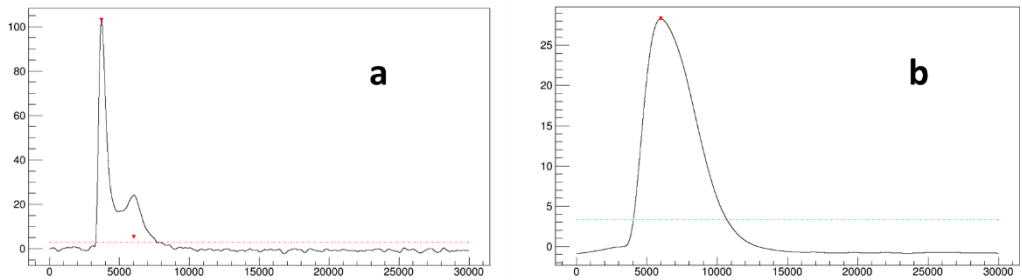

Fig. 8. A preamplifier waveform after the PSD, with different integration constants.

But at a higher count rate environment, the cathode with a thin metalized film is also a compromise instead.

On the other hand, the time difference contains information of the mobility of ions. It is a possible way for the velocity measurements of ions. Moreover, the ratio of the two peaks indicates the avalanche propagation around the wire. It offers a possibility to study the avalanche process and gas characteristics.

# Acknowledgments

The authors would like to acknowledge the funding support provided by the National Key R&D Program of China (Grant No. 2016YFA0401604).